\documentstyle[12pt]{article}
\newcommand{\newsection}[1]{
\addtocounter{section}{1}
\setcounter{equation}{0}
\setcounter{subsection}{0}
\addcontentsline{toc}{section}{\protect
\numberline{\arabic{section}}{{\rm #1}}}
\vglue .6cm
\pagebreak[3]
\noindent{\bf  \thesection. #1}\nopagebreak[4]\par\vskip .3cm}
\newcommand{\newsubsection}[1]{
\addtocounter{subsection}{1}
\addcontentsline{toc}{subsection}{\protect
\numberline{\arabic{section}.\arabic{subsection}}{#1}}
\vglue .4cm
\pagebreak[3]
\noindent{\it \thesubsection. #1}\nopagebreak[4]\par\vskip .3cm}

\renewcommand{\theequation}{\thesection.\arabic{equation}}

\newcommand{\ben}{\begin{enumerate}}
\newcommand{\een}{\end{enumerate}}
\newlength{\extraspace}
\newlength{\extraspaces}
\newcounter{dummy}
\newcommand{\bc}{\begin{center}}
\newcommand{\ec}{\end{center}}
\newcommand{\be}{\begin{equation}
\addtolength{\abovedisplayskip}{\extraspaces}
\addtolength{\belowdisplayskip}{\extraspaces}
\addtolength{\abovedisplayshortskip}{\extraspace}
\addtolength{\belowdisplayshortskip}{\extraspace}}
\newcommand{\ee}{\end{equation}}

\newcommand{\ba}{\begin{eqnarray}
\addtolength{\abovedisplayskip}{\extraspaces}
\addtolength{\belowdisplayskip}{\extraspaces}
\addtolength{\abovedisplayshortskip}{\extraspace}
\addtolength{\belowdisplayshortskip}{\extraspace}}
\newcommand{\ea}{\end{eqnarray}}

\newcommand{\ban}{\begin{eqnarray*}
\addtolength{\abovedisplayskip}{\extraspaces}
\addtolength{\belowdisplayskip}{\extraspaces}
\addtolength{\abovedisplayshortskip}{\extraspace}
\addtolength{\belowdisplayshortskip}{\extraspace}}
\newcommand{\ean}{\end{eqnarray*}}
\newcommand{\baa}{                         
\addtocounter{equation}{1}
\setcounter{dummy}{\value{equation}}
\setcounter{equation}{0}
\renewcommand{\theequation}{\thesection.\arabic{dummy}\alph{equation}}
\begin{eqnarray}
\addtolength{\abovedisplayskip}{\extraspaces}
\addtolength{\belowdisplayskip}{\extraspaces}
\addtolength{\abovedisplayshortskip}{\extraspace}
\addtolength{\belowdisplayshortskip}{\extraspace}}
\newcommand{\eaa}{                                       
\end{eqnarray}
\setcounter{equation}{\value{dummy}}
\renewcommand{\theequation}{\thesection.\arabic{equation}}}

\input epsf

\newcounter{fignum}
\newcounter{tabel}

\newcounter{tabnum}
\newcounter{xxx}
\newcommand{\bl}{\begin{list}{({\it\roman{xxx}})}{\usecounter{xxx}}}
\newcommand{\el}{\end{list}}
\newcommand{\ppt}[1]{{\partial \over \partial t}}            
\newcommand{\ppx}[1]{{\partial \over \partial x}}            
\newcommand{\pqt}[1]{{\partial^2 \over \partial t^2}}            
\newcommand{\pqx}[1]{{\partial^2  \over \partial x^2}}            
\renewcommand{\l}{\langle}

\def\a{\alpha} 
\def\b{\beta} 
\def\g{\gamma}

\def\l{\lambda}

\def\f{\phi}

\def\<{\langle}
\def\>{\rangle}

\newfont{\gothic}{eufm10 scaled\magstep1}

\newcounter{problems}

\begin{document}
\begin{titlepage}
\begin{flushleft}
\today
\end{flushleft}
\begin{center}
{\LARGE \bf Ground-state splitting around rotating Mini Blackholes}
\end{center}
\vskip 2cm
\begin{center}
\mbox{I.\ Sturm and F.M.C. Witte }\\
{\it
Institute of Theoretical Physics,\\
Department of Physics and Astronomy, Utrecht University\\
Leuvenlaan 4, 3584 CE, Utrecht\\
Netherlands}
\end{center}
\vskip 1.5cm
\begin{abstract}
In this letter we present the result of a spin-dependent groundstate-energy calculation for fermionic boundstates in the spacetime around a rotating blackhole. Using a slow rotation approximation and a minimax variational approach we find boundstate energies of 0 to 5 percent of the fermions flatspace restmass. The groundstate displays a spin-dependent splitting with an energy difference of about 10 percent of the binding energy. For a dilute gas of primordial mini blackholes with gravitationally bound electrons spin-flip transitions could possibly give rise to observable signatures in the observed soft X-ray spectrum for sources at cosmological distances.
\end{abstract}
\end{titlepage}
\newpage

\newsection{Introduction}
Studies in quantum gravity seek to elucidate how the notions of quantum physics can be reconciled with the description of gravity as given by General Relativity. A question that is at the forefront of this endeavour is how to view the quantum nature of Spacetime. We consider a different but related question. Two gravitationally interacting particles, with masses $M$ and $m$, in an asymptotically flat spacetime can be considered a most elementary system allowing an unproblematic outside observer. As a result its quantum treatment should not be hindered to much by fundamental questions regarding interpretation induced by quantum cosmological issues. So what can we learn from a study of the quantum states of a system of a pair of gravitationally interacting particles? 

We express the system's boundstate energies in units of $mc^2$, and use as length unit the Compton wavelength $\l_{c} = \frac{\hbar}{m c}$. The strength of the gravitational interaction between the two particles is then parametrized by $\a$,
\be
\a = \frac{G M}{\l_{c} c^2} \ ,
\ee
basically representing the ratio of gravitational- and  rest-energy of the bound particle. The treatment of systems with $\a >> 1$ will require considering the quantum field nature of the bound particle whereas for $\a \approx 0$ is a "Newtonian setting" where any quantum treatment of this system reduces to that of a hydrogenic atom. In between, say roughly between $\a = 0.1$ and $\a = 1$, is the range where ordinary quantum mechanics in Newtonian spacetime needs to be replaced by relativistic quantum mechanics in curved spacetime. The parameter which indicates whether a quantum treatment of gravity itself is neccesary is the ratio between the Planck length $\l_{p}$ and $\l_{c}$.
\be
\b = \frac{\l_{p}}{\l_{c}} = \sqrt{\a \frac{m}{M}} \ .
\ee
Obviously if $\b = 1$ the system's typical size is of the order of magnitude of the expected grainyness of quantum spacetime itself.
Notice that for moderate values of $\a$ the parameter $\b$ can come relatively close to $\b \approx 1$. So what type of systems are we talking about if $\a$ is close to but below unit value? If we consider a hypothetical primordial {\it mini blackhole} with $M \approx 10^{12}$ kg, binding an electron we have $\a \approx 0.5$ while the typical electron-Blackhole $\b$ will be vanishingly small at about $10^{-20}$. 

The structure of this letter is as follows. In section 2 we will briefly review several of the known results about boundstates of fermions in blackhole backgrounds. In section 3 we will present our results for the groundstate in the Kerr spacetime around a rotating blackhole as they were obtained in \cite{Sturm}. A detailed report 
on the computations neccesary to obtain these results will follow elswhere \cite{SturmWitte}.

\newsection{Boundstates around Blackholes}
Investigating bound state solutions in Blackhole systems is not a new endeavour. Doran, Lasenby \emph{et al.} showed \cite{Caceres Doran, Lasenby Doran Pritchard Caceres Dolan} that bound states, albeit instable due to the presence of the singularity, of a fermion around a black hole do exist in the Schwarzschild case, i.e. $Q=a=0$, $Q$ being the black hole net charge and $a$ the angular-momentum parameter of the black hole, defined as $\frac{S}{M}$. $S$ is the spin-momentum of the object and $M$ its mass. 

The Kerr-Newman metric has also been investigated on the
possibility of bound states. For certain values of the parameters
$M, a$ and $Q$, conclusions have been obtained about the
(im)possibility of stable bound states. It is, for instance, already
proven \cite{Finster Smoller Yau} that bound states with real energies do not occur in the so called Reissner-Nordstr\o m metric (Kerr-Newman, with $a=0$) nor \cite{Finster Kamran Smoller Yau} in the non-extreme case of Kerr-Newman geometry, i.e. $M^{2}>a^{2}+Q^{2}$. Schmid \cite{Schmid}, has shown that for the extreme Kerr case, i.e. $M^{2}=a^{2}$, bound states to the Dirac equation do exist in Kerr geometry. For each azimuthal quantum number $k$ and particular values of $a$, bound state solutions to the Dirac equation exist and the energy of these states is uniquely determined by $\omega=-\frac{kM}{2J}$. These bound states can only occur, when the product of the test particle and the black hole mass, $mM$, lies in the range $\frac{k}{2}<mM<\frac{k}{\sqrt{2}}$.

In this brief report we will focus our analysis on the groundstate energy in the so called linearized Kerr metric. We will linearize the Kerr metric in the parameter $a$, thus only looking at black holes with a relatively small spin. The reasons for studying the linearized Kerr metric are twofold. Firstly, the linearized Kerr geometry has only one horizon instead of the two from the full Kerr metric. In the extreme Kerr case there is also only one horizon, since the inner and outer horizon coincide. The established existence of bound states in this regime gives us a hint on the possibility for bound states when the black hole is only slowly rotating. Secondly, also in the Schwarzschild geometry instable bound states exist and the linearized Kerr metric can be considered to be a small perturbation of the Schwarzschild spacetime. The goal in the next section is to quantify the effect on he ground-state energy of this perturbation due to the Blackhole's angular-momentum in the regime of $\a < 1$.

\newsubsection{Variational principle}
When dealing with non-relativistic quantum mechanics, a powerful
tool in constructing eigenfunctions and eigenvalues of the
Schr\"{o}dinger equation, are variational methods. For relativistic Dirac Hamiltonians however, things aren't that simple. There is a difficulty in the variational determination of energy-eigenvalues called the variational collapse. It is due to the fact that the spectrum of the Dirac Hamiltonian is not bounded from below, allowing states with both positive and negative energy. Gazdy and Ladanyi \cite{Gazdy Ladanyi} showed that if the basis is
not chosen properly, the eigenvalues representing particle bound
states may decrease uncontrollably. A direct minimization of the binding energy takes $E_{b}$ to $-\infty$. In another context \cite{Datta}, stability of the solutions was ensured by projecting out the negative-energy states. There are methods which can help avoiding the variational collapse. The correct way as to tackle the problem is to choose the small-component parameters to maximize $E$ and to choose the large-component parameters to minimize $E$. A discussion on this is given in \cite{Wood Grant Wilson} and \cite{Talman}. The appropriate formulation for the Dirac equation particle ground state is therefore
\ba
\label{minimaxequation}
E={min}_{g}\Biggl[{max}_{f}\frac{\langle\Phi|{\bf
H}|\Phi\rangle}{\langle\Phi|\Phi\rangle}\Biggr]\;,
\ea
with $g(r)$ and $f(r)$ respectively the large and small components
of the spinor $\Phi(\vec{r})$. The trial function $\Phi(\vec{r})$ of (\ref{minimaxequation}) can
be expanded in a finite function basis as follows
\ba
\label{1.38}
\Phi(\vec{r})=\sum_{j=1}^{n}\left(\begin{array}{cc}b_{j}g_{j}(r)\\a_{j}f_{j}(r)\end{array}\right)\;.
\ea
The Dirac equation in Blackhole backgrounds has a number of singular points. To find a proper set of trial functions the analytic behaviour of the solutions to the Dirac equation near these singular points must be determined.

\newsubsection{The Schwarzschild spacetime}
The Schwarzschild solution is, according to Birkhoff's theorem \cite{Birkhoff}, the most general static, spherically symmetric, vacuum solution to Einstein's field equations. It describes the gravitational field outside a spherically symmetric non rotating mass, without charge. The Schwarzschild metric is given by
\ba\label{PG-metric}
ds^{2}=&-\Bigl(1-\frac{2M}{r}\Bigr)dt^{2}+2\Bigl(\frac{2M}{r}\Bigr)^{1/2}drdt+dr^{2}+r^{2}d\Omega^{2}\nonumber\\
=&-dt^{2}+\Bigl(dr+\Bigl(\frac{2M}{r}\Bigr)^{1/2}dt\Bigr)^{2}+r^{2}d\Omega^{2}\;.
\ea
in the {Painlev\'{e}-Gullstrand} form. We used this form to compute the groundstate energies, the results can be shown to be gauge-independent. On a gravitational background the Dirac equation has the form
\ba
i\gamma^{\mu}D_{\mu}(x^{\nu})\Psi=m\Psi\;.
\ea
where $D_{\mu}$ is the covariant derivative on spinors containing the spin-connection. The computation of the spin-connection requires the use of the {tetrad} or {vierbein} formalism. The spin connection for the Painlev\'{e}-Gullstrand gauge, of which the naturally associated tetrad is very simple \cite{Dolan}, can be put into the following form
\ba
g^{\mu}\frac{i}{2}\omega_{\mu}^{ab}\Sigma_{ab}=-\frac{3}{4r}\sqrt{\frac{2M}{r}}\gamma_{0}\;.
\ea
Expressing all lengths in terms of the Compton wavelength simplifies the resulting interaction hamiltonian to
\ba
H_{1}(r)=i\sqrt{\frac{2\alpha}{r}}(\partial_{r}+\frac{3}{4r}).
\ea
This system possesses 3 singular points. One at infinity, one at the origin and one at $r=2\alpha$. For each singularity we rewrite our equations into a form where the singular part of the equation is separated from the non-singular part of the equation. The non singular part is thus analytic at that singularity, making it possible to use Frobenius' method to expand about the singularity. From this analysis we obtain the general form of the trial functions
\ba
g(r),\;f(r)\propto R(r)e^{-\sqrt{1-E^{2}}r}e^{2i\sqrt{2\alpha
r}E}r^{-3/4}\;.
\ea
Using this method with an expansion up to $j=4$ we calculate the binding energy of the fermion at a representative $\a = 0.1$ to be equal to
\ba
E_{b}=(0.995-1) mc^{2}= -0.005 mc^{2} \ .
\ea
If we are dealing with a bound electron $E_{b}$ is thus $-2.6$ KeV. To get an impression of this number compare it to the groundstate of hydrogen $-13.7$ eV. Our results are in agreement with known results \cite{Caceres Doran} for $\a$ in the range between 0 and 0.4. 
The computation of $E_{b}$ at higher values of $\alpha$ is straightforward but requires a substantially larger number of terms in the expansion. In \cite{Caceres Doran} the full $\a$ dependence has been studied. For our purposes settling for $\a < 0.4$ is sufficient as we are mainly interested in the effects of Blackhole angular momentum here.

\newsection{Results on the Kerr spacetime}
The metric for a black hole with mass $M$, spin $S$ is the Kerr metric:
\ba
ds^{2}=-dt^{2}+\frac{\rho^{2}}{\Delta}dr^{2}+\rho^{2}d\theta^{2}+(r^{2}+a^{2})\sin^{2}\theta
\;d\phi^{2}+\frac{2Mr}{\rho^{2}}(a \sin^{2}\theta
\;d\phi-dt)^{2}\;,
\ea
where $a = \frac{S}{M}$. The metric coefficients will be expanded up to first order
in $a$
\ba
g_{\mu\nu}(a,r,\theta)=g_{\mu \nu}(0,r,\theta)+\partial_{a}g_{\mu
\nu}(a,r,\theta)|_{a=0}\cdot a.
\ea
The zeroth order term in the expansion gives us back the
Schwarzschild metric coefficients. The first order term in $a$ may
thus be seen as a perturbation on the Schwarzschild metric. To compute the spin-connection in the Kerr geometry we chose the Kinnersley null tetrad proposed by Chandrasekhar \cite{Chandrasekhar}. After calculation of the spin-connection, we arrive at the Dirac equation in Kerr geometry:
\ba\label{Dirac Kerr}
({\bf R}+{\bf A})\Psi(x^{\mu})=0\;,
\ea
with
\ba
{\bf R}:=\left(\begin{array}{cccc}
imr&0&\sqrt{\Delta}D_{+}&0\\
0&-imr&0&\sqrt{\Delta}D_{-}\\
\sqrt{\Delta}D_{-}&0&-imr&0\\
0&\sqrt{\Delta}D_{+}&0&imr \end{array}\right)\;,
\ea
and
\ba
{\bf A}:=\left(\begin{array}{cccc}
-am\cos\theta&0&0&L_{+}\\
0&am\cos\theta &-L_{-}&0\\
0&L_{+}&-am\cos\theta&0\\
-L_{-}&0&0&am\cos\theta \end{array}\right)\;.
\ea
The differential operators $L_{\pm}$ and $D_{\pm}$ are given by
\ba
D_{\pm}&=\partial_{r}\mp\frac{1}{\Delta}\Bigl[(r^{2}+a^{2})\partial_{t}+a\partial_{\phi}\Bigr]\;,
\nonumber\\  L_{\pm}&=\partial_{\theta}+\frac{\cot\theta}{2}\mp
i\Bigl[a\sin\theta\partial_{t}+\frac{1}{\sin\theta}\partial_{\phi}\Bigr]\;.
\ea
For the Dirac spinor we can now use the following Ansatz
\ba
\Psi(t,r,\theta,\phi)=e^{-iE
t}e^{-ik\phi}\left(\begin{array}{cccc}g(r)\chi_{1}(\theta)\\f(r)\chi_{2}(\theta)\\f(r)\chi_{1}(\theta)\\g(r)\chi_{2}(\theta)
\end{array}\right),\ea
where $k\in \{ \pm \frac{1}{2},\pm \frac{3}{2}\}$ is a half
integer often referred to as the magnetic quantum number and $E$ is real. We introduce a real-valued separation parameter $\lambda$  to obtain the two equations
\ba
{\bf R}\Psi=\lambda\Psi, \;\;\;{\bf A}\Psi=-\lambda\Psi\;,
\ea
to separate the radial and the angular part. We solved the angular equation first to find an expression for the separation parameter $\lambda$ as a function of the energy $E$ in the small $\a$ regime.This case can be solved with the help of a perturbative procedure described
in \cite{Chakrabarti}. The main idea is to write the solutions as
an expansion in the so-called spin-weigthed spherical harmonics $_{\frac{1}{2}}S_{lk}$. We have:
\begin{eqnarray}
_{\frac{1}{2}}S_{lk} &=&\sum_{l_{'}}C_{l^{'}}^{+}\:
_{\frac{1}{2}}Y_{lk}\;,\\
_{-\frac{1}{2}}S_{lk}&=&\sum_{l_{'}}C_{l^{'}}^{-}\:
_{-\frac{1}{2}}Y_{lk}\;,
\end{eqnarray}
with
\ba
C_{l^{'}}^{+}=(-1)^{(l'-\!l)}C_{l^{'}}^{-}\;,
\ea
and to treat the terms
\ba\label{3.55}
\bigl[a^{2}(E^{2}-m^{2})\cos^{2}\theta-aE\cos\theta\bigr]\chi_{2}\;\;{and}\;\;am\sin\theta \chi_{1}\;,
\ea
as perturbations. The matrix element with respect to these terms
can be written in the form
\ba\label{3.56}
H_{l',l}=-aE _{1}C^{l}_{l'}+am(-1)^{(l'-\!l)}
\:_{1}D^{l}_{l'}+a^{2}E^{2}\Bigl(1-\frac{m^{2}}{E^{2}}\Bigr)
_{2}C^{l}_{l'}\;,
\ea
with
\ba
_{1}C^{l}_{l'}=&\int\:_{-\frac{1}{2}}Y^{*}_{l'k-\frac{1}{2}}Y_{lk}
\cos\theta d\Omega \nonumber\\
=&\Bigl(\frac{2l+1}{2l'+1}\Bigr)^{\frac{1}{2}}\langle
l\;1\;k\;0\;|\;l'\;k\rangle \langle
l\;1\;\frac{1}{2}\;0\;|\;l'\;\frac{1}{2}\rangle\;,
\ea
\ba
_{1}D^{l}_{l'}=&\int \:_{-\frac{1}{2}}Y^{*}_{l'k
\frac{1}{2}}Y_{lk}
\sin\theta d\Omega \nonumber\\
=&-\Bigl(2\frac{2l+1}{2l'+1}\Bigr)^{\frac{1}{2}}\langle
l\;1\;k\;0\;|\;l'\;k\rangle\langle
l\;1\;-\frac{1}{2}\;1\;|\;l'\;\frac{1}{2}\rangle\;,
\ea
and
\ba
_{2}C^{l}_{l'}=&\int \:_{-\frac{1}{2}}Y^{*}_{l'k
-\frac{1}{2}}Y_{lk}
\cos^{2}\theta d\Omega \nonumber\\
=&\frac{1}{3}\delta_{l'l}+\frac{2}{3}\Bigl(\frac{2l+1}{2l'+1}\Bigr)^{\frac{1}{2}}\langle
l\;2\;k\;0\;|\;l'\;k\rangle \langle
l\;2\;-\frac{1}{2}\;0\;|\;l'\;\frac{1}{2}\rangle\;.
\ea

The $\bigl<l_{1}\;l_{2}\;k_{1}\;k_{2}\mid l\;k\bigr>$ are the
Clebsch-Gordan coefficients, as defined in i.e. \cite{Abramowitz
Stegun}. Finally, with the matrix elements as defined in (\ref{3.56}), we can use the standard method of Brillouin and Wigner to obtain the terms in the perturbative expansion of $\lambda^{2}$. The ground state is given by the combination $\{l,k\}=\{\frac{1}{2},+\frac{1}{2}\}$ and some simple but time consuming algebra gives the following result
\ba\label{lambda equation}
\lambda^{2}(a,m,E)=1+\frac{2}{3}aE-2a^{2}\Bigl(-1+\frac{2}{27}\bigl(1+\frac{m}{E}\bigr)^{2}\Bigr)E^{2}+{\mathcal{O}}((aE)^{3})\;.
\ea
We find for the eigenfunction
\ba\label{angular part kerr}
_{\frac{1}{2}}S_{\frac{1}{2}\frac{1}{2}}&=\frac{i}{\sqrt{2\pi}}\sin\Bigl(\frac{\theta}{2}\Bigl)e^{i\phi/2}+\frac{1}{9}\Bigl(\frac{m}{E}+1\Bigr)aE\frac{i}{\sqrt{2\pi}}\sin\Bigl(\frac{\theta}{2}\Bigl)e^{i\phi/2}(1+3\cos(\theta))\;.\nonumber\\
&=\frac{i}{\sqrt{2\pi}}\sin\Bigl(\frac{\theta}{2}\Bigl)e^{i\phi/2}\Bigl[1+\frac{1}{9}\Bigl(\frac{m}{E}+1\Bigr)(1+3\cos(\theta))aE\Bigr]\;.
\ea
There is no need to compute the state with reversed spin as its energy eigenvalue can be obtained by simply substituting $a \rightarrow -a$ from symmetry considerations. Using the notation $\l$ for the linear approximation to $\l(am, aE)$ and analyzing the form of the trial functions that properly suits the singular points of the Dirac equation we obtain
\ba
g(r),f(r)\propto r^{\pm\frac{iak}{2\alpha}}
(r-2\alpha)^{\pm\frac{i}{2}\bigl(\frac{ak}{\alpha}+4\alpha
E\bigr)}e^{-\sqrt{m^{2}-E^{2}}r}r^{-\frac{E}{m}\lambda}\;.
\ea
In these coordinates the Dirac Hamiltonian reads
\ba
H=\frac{\Delta}{r^{2}+a^{2}}\left(\begin{array}{cc}
i\partial_{r}-\frac{ka}{\Delta}&\frac{1}{\sqrt{\Delta}}(-mr-i\lambda)\\\frac{1}{\sqrt{\Delta}}(-mr+i\lambda)&-i
\partial_{r}-\frac{ka}{\Delta}\end{array}\right)\;.
\ea
The term in front of the matrix is the inverse of the measure of the Kerr metric. So when calculating integrals that term cancels the
measure term in the integral. Formally the the minimax procedure requires the Hamiltonian to be hermitian. In the Schwarzschild case the singularity at the origin gives rise to a small non-hermiticity. In the Kerr case, as we are not interested in determining the decay time of the boundstates, we restrict the integrations to spacetime outside the horizon where the given Hamiltonian is hermitian. All the above is now input for the variational determination of the groundstate energy of a fermion in the (linearised) Kerr spacetime.

\newsubsection{Level-splitting in the Kerr groundstate}
The first thing to do, is to check what value for the
ground state energy we find if we set $a$ equal to zero. This
means a Kerr black hole without rotation, thus resulting in the
Schwarzschild metric. In chapter 2 we found the ground state value
of $0.9955 mc^2$. For the Kerr case with $a=0$ we find the value $0.9945 mc^2$. The two values are sufficiently close together to be considered reliable. The fact that they aren't exactly the same is because we for the Kerr metric we did not include the spacetime inside the horizon in our variational calculation. We checked how the value of $\alpha$ influences the groundstate energy. For low values of the parameter $a<1$, the energy of the groundstate of the system falls off with increasing $\alpha$. This is also the case in the Schwarzschild background. But what is of interest here interesting is that there is a difference in the value of the groundstate for negative or positive values of the spin parameter $a$. When the orientation of the angular momentum of the black hole is antiparallel to that of the fermion the binding energy is less. A maximum value of for the energy-split is found at $\alpha=0.20$ and rapidly vanishes towards $\a \rightarrow 0$. For $\a > 0.2$ there is a slow decrease. Typically the maximal level-splitting is given by
\be
\Delta E_{b} \approx 0.3 a mc^2 \ .
\ee 
As a result a spin-flipping transition of a bound electron near a slowly rotating primordial blackhole should be typically in the 1 keV range. The corresponding electromagnetic radiation would be in the soft X-ray range. It is conceivable that a dilute gas of primordial mini Blackholes has remained as a remnant after the Big Bang. The availabillity of the spin-flip process for bound electrons would allow for extra absorption of soft X-rays emmited by X-ray sources at large distances. Work to establish the precise observable, spectral and polarization characteristics of such a signal is currently in progress.

\end{document}